\documentclass
[aps,prb,twocolumn,floatfix,english,showpacs,10pt,superscriptaddress]{revtex4-2}
\usepackage{graphicx}
\usepackage{booktabs}
\usepackage{amsmath}
\usepackage{physics}
\usepackage{amssymb}
\usepackage{colordvi}
\usepackage{verbatim}
\usepackage{xcolor}
\usepackage{mathrsfs}
\usepackage{epsfig}
\usepackage{lipsum}
\usepackage{amsfonts}
\usepackage{makecell}
\usepackage{multirow}
\usepackage{bm}

\usepackage[unicode=true, breaklinks=false, pdfborder={0 0 1}, backref=false,
colorlinks=true, linkcolor=blue, urlcolor=blue, citecolor=blue]{hyperref}%
\providecommand{\U}[1]{\protect\rule{.1in}{.1in}}
\setcitestyle{numbers,square}

\begin{document}

\title{Spin quenching and transport by hidden Dzyaloshinskii-Moriya interactions}

\author{Xiyin Ye}
\affiliation{School of Physics, Huazhong University of Science and Technology, Wuhan 430074, China}

\author{Qirui Cui}
\affiliation{Department of Applied Physics, School of Engineering Sciences, KTH Royal Institute of Technology, AlbaNova University Center Stockholm SE-10691, Sweden}

\author{Weiwei Lin}
\affiliation{Key Laboratory of Quantum Materials and Devices of Ministry of Education, Southeast University, Nanjing 211189, China}
\affiliation{School of Physics, Southeast University, Nanjing 211189, China}

\author{Tao Yu}
\email{taoyuphy@hust.edu.cn}
\affiliation{School of Physics, Huazhong University of Science and Technology, Wuhan 430074, China}
\affiliation{Key Laboratory of Quantum Materials and Devices of Ministry of Education, Southeast University, Nanjing 211189, China}
\date{\today}

\begin{abstract}

Explicit interactions, \textit{e.g.}, dipolar and exchange couplings, usually govern magnetization dynamics. Some interactions may be hidden from the global crystal symmetry. We report that in a large class of \textit{uniaxial} antiferromagnets, a \textit{hidden} Dzyaloshinskii-Moriya interaction with retaining global inversion symmetry quenches the spin of magnon along the N\'eel vector ${\bf n}$, thus forbidding its angular-momentum flow. Some magnon spins, termed ``nodal" and ``corner" spins, survive when they distribute \textit{singularly} at the hot spots, i.e., high-symmetric degeneracy points in the Brillouin zone, and are protected by crystal symmetries. The biased magnetic field along ${\bf n}$ broadens such distributions, allowing bulk spin transport with unique signatures in the magnetic field and temperature dependencies. This explains recent experiments and highlights the role of hidden interaction. 

\end{abstract}

\maketitle

\section{Introduction}

Orbital quenching by crystal fields in solids is common~\cite{Kittel,Orbital}. The spins, on the other hand, are not quenched in most ferromagnets and antiferromagnets, allowing the transport of angular momentum in insulators by magnons~\cite{FM1,FM2,FM3,FM4,FM5,AFM1,AFM2,AFM3}. The long-distance spin transport in uniaxial  antiferromagnets~\cite{easy_axis1,easy_axis2,easy_axis6,easy_axis7,easy_axis8,easy_axis9,easy_axis10,easy_axis11,easy_axis12,easy_axis13,YFeO,easy_axis3,easy_axis4,easy_axis5,easy_axis14,easy_axis15,easy_axis16,easy_axis17,YFeO2,W.Lin,swapping1,swapping2} is enabled by magnons of integer spins. Exceptions are easy-plane antiferromagnets in which the modes are linearly polarized or the spins are ``quenched" by the anisotropies~\cite{easy_plane0,easy_plane0_,easy_plane1,easy_plane2,easy_plane3,easy_plane4,easy_plane5}. Nevertheless, recent experiments show evidence that the spins (along the N\'eel vector) are quenched in uniaxial antiferromagnets when canted by small  angles~\cite{YFeO,W.Lin,swapping1,swapping2,easy_axis3,easy_axis4} such that spin transport is suppressed when biased by low magnetic fields. 
Here, we report the proper canted magnetic ground state is the source of ``spin quenching" in a large class of uniaxial antiferromagnets.  The local (hidden) Dzyaloshinskii-Moriya interaction (DMI) acts as an intrinsic mechanism leading to the canted magnetic configuration and unique spin configuration of magnons.

\begin{figure}[htb]	
\includegraphics[width=0.47\textwidth,trim=0.6cm 0cm 0cm 0.1cm]{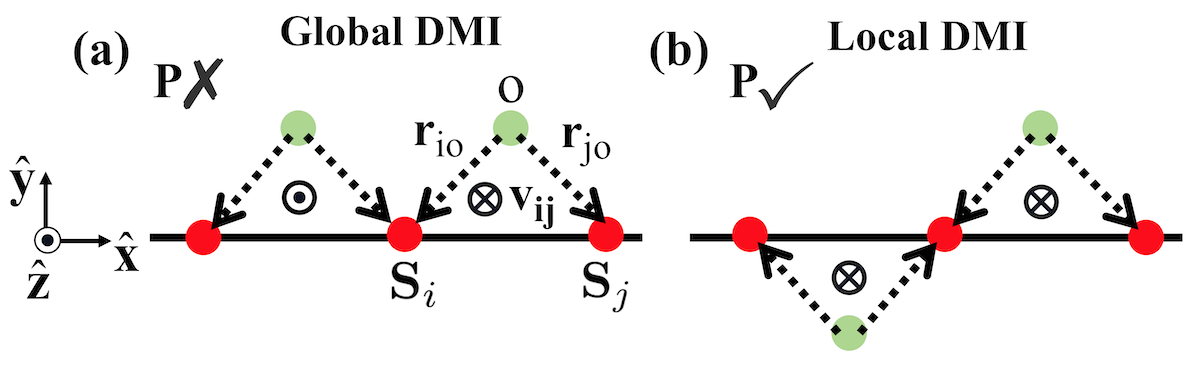}
\caption{Global \textit{vs}. local (hidden) DMI.}
\label{DMI}
\end{figure}

DMI is an asymmetric exchange interaction between magnetic moments due to the spin-orbit coupling~\cite{1,2,3,s1}, which is widely measured in non-centrosymmetric magnets or magnet$|$heavy metal interfaces~\cite{s2,s3,s4} and stabilizes noncollinear magnetic orders such as chiral domain walls, skyrmions, and periodic canted spins~\cite{4,5,6}. The emergence of DMI between magnetic ions requires breaking inversion symmetry, as shown in Fig.~\ref{DMI}(a). Interestingly, a local (hidden) DMI can also be induced by local breaking of inversion symmetry as in Fig.~\ref{DMI}(b) in centrosymmetric magnets~\cite{7,8}, leading to noncollinear spin configurations where the sublattice displays opposite chirality. A large class of uniaxial antiferromagnets, \textit{e.g.}, $R$FeO$_3$ ($R=$ rare earth)~\cite{RFeO1,RFeO2,RFeO3,RFeO4,RFeO5}, $R$MnO$_3$~\cite{RMnO}, Ca$_2$RuO$_4$~\cite{CaRuO1,CaRuO2},  BaCoS$_2$~\cite{7,BaCoS1,BaCoS2}, and bilayer MnBi$_2$Se$_2$Te$_2$~\cite{MBST}, hold hidden DMI and allow long-distance spin transport~\cite{YFeO,YFeO2,W.Lin,swapping1,swapping2}.
To the best of our knowledge, understanding the role of such local DMI on the spin transport of magnons remains wanting. 
In the canted antiferromagnets YFeO$_3$ with local DMI, the absence of spin transport without magnetic field and long-range magnon transport enabled by the magnetic field~\cite{YFeO} indicate a distinct spin-transport mechanism from those in easy-axis and easy-plane antiferromagnets, noting in easy-axis antiferromagnets magnons carry spin angular momentum via intrinsic circularly polarized modes~\cite{easy_axis12} while in easy-plane antiferromagnets spin transport is supported by pairing two linearly polarized modes~\cite{easy_plane2,easy_plane3}.

In this work, we find that the hidden DMI, via tilting the ground-state magnetization, generally quenches the spin of magnons along the N\'eel vector in uniaxial antiferromagnets.
But there are also momentum-space ``hotspots", i.e., high-symmetric degeneracy points, where the hidden DMI is absent, at which the magnon spins survive and distribute singularly in the Brillouin zone (BZ). Crucially, these spins are protected by the crystal symmetries. 
The applied magnetic field along the N\'eel vector broadens their distribution, allowing bulk spin transport at low magnetic fields. We predict a peak in the spin conductivity in the temperature dependence and find a scaling law to the applied magnetic field that agrees with the experiment~\cite{YFeO}.

\section{Simple model}

\begin{figure}[htb]	
\includegraphics[width=0.42\textwidth,trim=0.6cm 0cm 0cm 0.1cm]{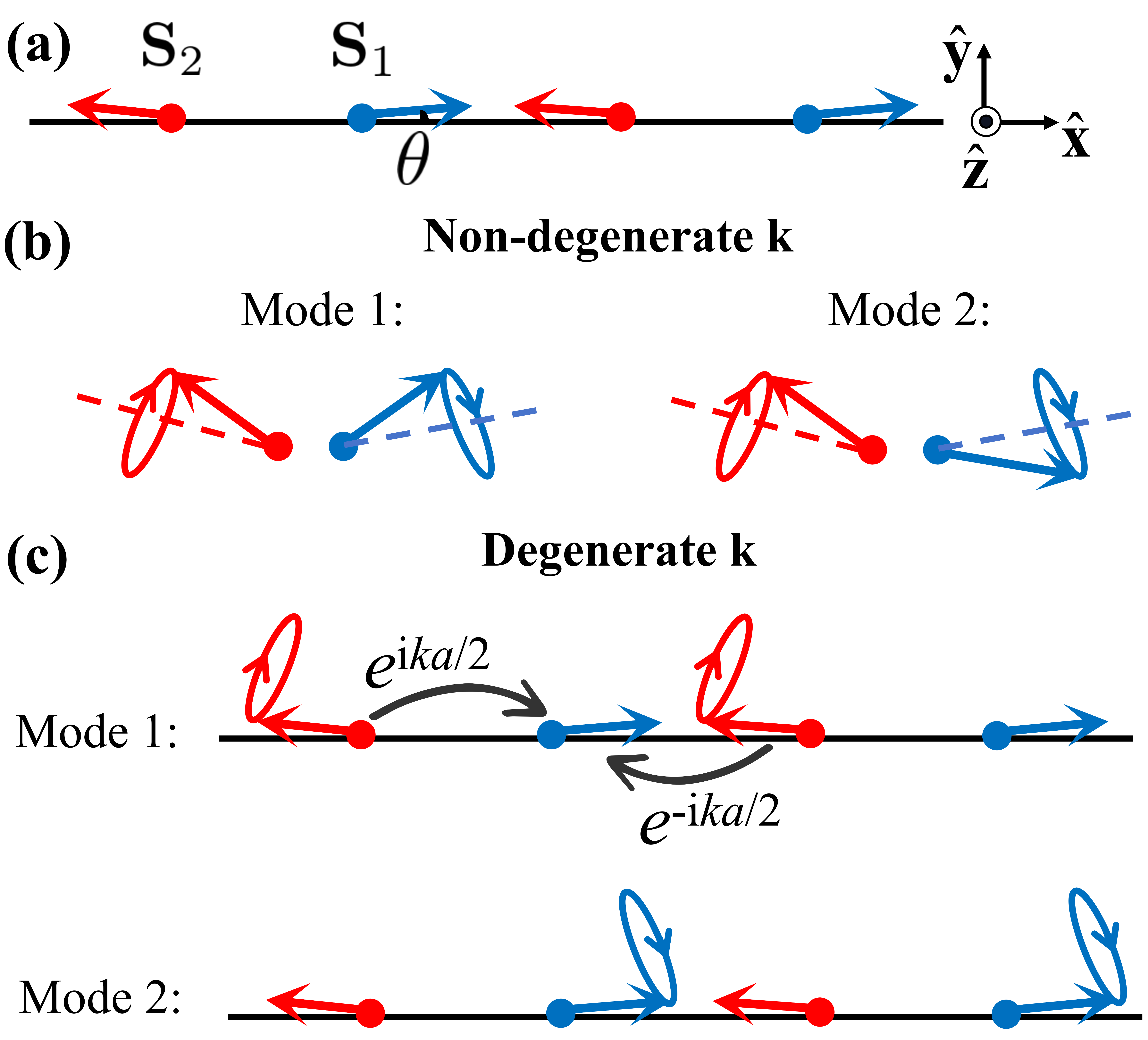}
\caption{Spins of magnon in uniaxial antiferromagnets canted by the hidden DMI. }
\label{1D}
\end{figure}

To show the role of hidden DMI on the spin quenching, we first consider a simple uniaxial antiferromagnetic chain as in Fig.~\ref{1D}(a), where the hidden DMI with strength $D$ breaks the local inversion symmetry of nearest-neighboring sites but maintains the global inversion symmetry [Fig.~\ref{DMI}(b)]. Two nearest-neighboring spins couple via the antiferromagnetic exchange interaction with coupling $J>0$ and are aligned along the $\hat{\bf x}$-axis by the easy-axis anisotropy $K_x<0$, governed by the Hamiltonian
\begin{align}
\hat{H}&=J\sum_{\langle i,j\rangle} \hat{\bf S}_{i} \cdot \hat{\bf S}_{j}-D\sum_{\langle i,j\rangle}{\bf v}_{ij}\cdot(\hat{\bf S}_{i} \times \hat{\bf S}_{j})+K_x\sum_{i}(\hat{S}_{i}^x)^2,
    \label{1D_H}
\end{align}
where $\langle \cdots \rangle$ represents the summation over the nearest-neighboring sites $\{i,j\}$. The anti-symmetric DM unit vectors ${\bf v}_{ij}=({\hat{\bf r}_i}-\hat{\bf r}_O)\times (\hat{\bf r}_O-\hat{\bf r}_j)=-{\bf v}_{ji}$ with $O$ representing a non-magnetic atom as in Fig.~\ref{DMI}~\cite{easy_axis2,DM,RFeO3,1,2,3,s1}. The local DMI ${\bf v}_{ij}={\hat{\bf z}}$ differs from the global one ${\bf v}_{ij}=\pm{\hat{\bf z}}$ (Fig.~\ref{DMI}), which stabilizes a canted ground-state configuration with $\hat{\bf S}_1=(\cos \theta, \sin \theta)$ and $\hat{\bf S}_2=(-\cos \theta, \sin \theta)$ in one magnetic unit cell, minimizing the free energy (\ref{1D_H}) with $2\sqrt{(2J-K_x)^2+4D^2} \sin(2\theta-\phi)=0$, where $\phi=\arctan \left[2D/(2J-K_x)\right]$. The canted angle $\theta=\phi/2\approx D/(2J-K_x)$ along $\hat{\bf y}$ is small when $D \ll J$.

With bosonic operators $(\hat{a}_{1,k_x},\hat{a}_{2,k_x},\hat{a}^{\dagger}_{1,-k_x},\hat{a}^{\dagger}_{2,-k_x})^T$ in the wave-vector ${\bf k}$ space as the basis, the Hamiltonian \eqref{1D_H} is linearized to be 
\begin{align}
    H(k_x)=\frac{S}{2}\begin{pmatrix}
        A & C_{+}(k_x) & B & C_{-}(k_x) \\
        C_{+}(k_x) & A & C_{-}(k_x) & B \\
        B & C_{-}(k_x) & A & C_{+}(k_x) \\
        C_{-}(k_x) & B & C_{+}(k_x) & A
    \end{pmatrix},
    \label{1D_H2}
\end{align}
where $S=|{\bf S}|$, $A=2A_0+2K_x(\sin^2\theta-2\cos^2\theta)$ with $A_0=D\sin2\theta+J\cos2\theta$, $B=K_x\sin^2\theta$,  and $C_{\pm}(k_x)=\pm(J \mp A_0)\cos(k_xa/2)$ with $a$ being the lattice constant of the chain. $H(k_x)=H(-k_x)$ due to the global inversion symmetry. We present the details of the Bogoliubov transformation in Appendix~\ref{appendix_A}.
The dispersion of the two modes 
\begin{align}
    \hbar \omega_1(k_x)&=({S}/{2})\sqrt{(A-C_{+}(k_x))^2-(B-C_{-}(k_x))^2}, \nonumber \\
    \hbar \omega_2(k_x)&=({S}/{2})\sqrt{(A+C_{+}(k_x))^2-(B+C_{-}(k_x))^2},\nonumber
\end{align}
are two positive eigenvalues of $\eta_0 H(k_x)$, where the metric $\eta_0={\rm diag}\{I_{2\times 2},-I_{2\times 2}\}$.  
We adopt a hyperbolic parametrization in terms of parameters $\{\alpha,\beta,m,l\}$: $A-C_{+}=m\cosh{\alpha}$, $A+C_{+}=l\cosh{\beta}$, $B-C_{-}=m\sinh{\alpha}$, and $B+C_{-}=l\sinh{\beta}$ such that $\hbar \omega_{1}=Sm/2$ and $\hbar \omega_{2}=Sl/2$. When $k_x=\pm \pi/a$, the modes are degenerate since $\cos(k_xa/2)=0$ implies $C_{\pm}=0$. Otherwise, the modes are not degenerate.

When $\omega_1\ne \omega_2$, the corresponding eigenvectors of $\eta_0 H(k_x)$ are
\begin{align}
    \Psi_{1+}&=\frac{\sqrt{2}}{2}\left(
\begin{array}{c}
\cosh{\frac{\alpha}{2}} \\
-\cosh{\frac{\alpha}{2}} \\
-\sinh{\frac{\alpha}{2}}\\
\sinh{\frac{\alpha}{2}}
\end{array}
\right),~~~~
\Psi_{2+}=\frac{\sqrt{2}}{2}\left(
\begin{array}{c}
-\cosh{\frac{\beta}{2}} \\
-\cosh{\frac{\beta}{2}} \\
\sinh{\frac{\beta}{2}}\\
\sinh{\frac{\beta}{2}}
\end{array}
\right), \nonumber\\ 
\Psi_{1-}&=\frac{\sqrt{2}}{2}\left(
\begin{array}{c}
\sinh{\frac{\alpha}{2}} \\
-\sinh{\frac{\alpha}{2}} \\
-\cosh{\frac{\alpha}{2}}\\
\cosh{\frac{\alpha}{2}}
\end{array}
\right),~~~~
\Psi_{2-}=\frac{\sqrt{2}}{2}\left(
\begin{array}{c}
-\sinh{\frac{\beta}{2}} \\
-\sinh{\frac{\beta}{2}} \\
\cosh{\frac{\beta}{2}}\\
\cosh{\frac{\beta}{2}}
\end{array}
\right).
\label{State1}
\end{align} 
With these eigenmodes the magnon spins with $\omega_1\ne \omega_2$ are calculated as $S_1^x=S_2^x=0$ according to Eq.~\eqref{eq_S} in the Appendix~\ref{appendix_A}.
Therefore, these magnon spins vanish due to the spin-quenching effect.

On the other hand, when the two bands are degenerate with $\omega_1=\omega_2$, 
the eigenstates of $\eta_0 H(k_x)$ and the magnon spins are ambiguous to define since the superposition of two degenerate eigenstates is still the eigenstates. 
To this end, we consider a small magnetic field applied along the N{\'e}el vector to slightly break the mode degeneracy. This allows the magnon spin to be obtained at the degenerate points by taking the zero-field limit.
As illustrated in Fig.~\ref{1D}(a), a small magnetic field $H_x\hat{\bf x}$ is applied along the N\'eel vector $\hat{\bf x}$-direction, which breaks the mode degeneracy at $k_x=\pm \pi /a$ while ensuring a negligible rotation of the N{\'e}el vector. 
Including the Zeeman coupling, 
the Hamiltonian with $k_x=\pm \pi /a$ under the basis $(\hat{a}_{1,k_x},\hat{a}^{\dagger}_{1,-k_x})^T$ and $(\hat{a}_{2,k_x},\hat{a}^{\dagger}_{2,-k_x})^T$ can be decomposed into two diagonal blocks
\begin{align}
    \hat{H}_{1(2)}=\frac{S}{2}\begin{pmatrix}
        A\mp H_{\rm Z} & B \\
        B & A\mp H_{\rm Z} \\
        \end{pmatrix},
\end{align}
where $H_{\rm Z}=g\mu_B \mu_0 H_x \cos\theta/S$ with Land{\'e} factor $g$ of electrons. 
The frequency of the two modes 
$\hbar \omega_{1(2)}=({S}/{2})\sqrt{(A\mp H_{\rm Z})^2-B^2}$ is no longer degenerate. 
In terms of a hyperbolic parametrization $\{\alpha',\beta',m',l'\}$ with  
\begin{align}
    &\begin{cases} 
    A-H_{\rm Z}=m'\cosh{\alpha'} \\
    B=m'\sinh{\alpha'}
 \end{cases},~~
 \begin{cases}
        A+H_{\rm Z}=l'\cosh{\beta'} \\
        B=l'\sinh{\beta'}
    \end{cases},
\end{align}
the corresponding eigenstates
\begin{align}
    \Psi_{1,+}&=\left(
    \begin{array}{c}
    -\cosh{\frac{\alpha'}{2}} \\
    \sinh{\frac{\alpha'}{2}}\\
    \end{array}
    \right),~~~~
    \Psi_{2,+}=\left(
    \begin{array}{c}
    -\cosh{\frac{\beta'}{2}} \\
    \sinh{\frac{\beta'}{2}}
    \end{array}
    \right),\nonumber\\
    \Psi_{1,-}&=\left(
    \begin{array}{c}
    -\sinh{\frac{\alpha'}{2}} \\
    \cosh{\frac{\alpha'}{2}}\\
    \end{array}
    \right),~~~~
    \Psi_{2,-}=\left(
    \begin{array}{c}
    -\sinh{\frac{\beta'}{2}} \\
    \cosh{\frac{\beta'}{2}}
    \end{array}
    \right), 
\label{State2}
\end{align}
lead to the magnon spins along the N\'eel vector of modes ``$1$" and ``$2$" 
\begin{align}
   S^x_1&=-\cos\theta \left(\cosh^2{\frac{\alpha'}{2}}+\sinh^2{\frac{\alpha'}{2}}\right), \nonumber \\
   S^x_2&=\cos\theta \left(\cosh^2{\frac{\beta'}{2}}+\sinh^2{\frac{\beta'}{2}}\right).
\end{align}
These spins do not vanish under small magnetic fields. When $H_x \rightarrow 0$, $(\alpha',\beta')\rightarrow \alpha$ such that the modes become degenerate. We obtain the magnon spin along the N\'eel vector without ambiguity in this limit.

Thereby, we find the magnon spins of modes $``1"$ and $``2"$ along the N\'eel vector $\hat{\bf x}$~\cite{N.Okuma}:
\begin{align}
    S^x_1&=\left\{\begin{array}{cc}
    0,\hspace{4.3cm}\omega_1\ne \omega_2\\
    -\cos\theta \left(\cosh^2{\frac{\alpha}{2}}+\sinh^2{\frac{\alpha}{2}}\right),~~~\omega_1=\omega_2
    \end{array}\right., \nonumber \\
    S^x_2&=\left\{\begin{array}{cc}
    0,\hspace{4.3cm}\omega_1\ne \omega_2\\
    \cos\theta \left(\cosh^2{\frac{\alpha}{2}}+\sinh^2{\frac{\alpha}{2}}\right),~~~~~\omega_1=\omega_2
    \end{array}\right.,
\end{align}
which vanish at the non-degenerate points $\omega_1\neq \omega_2$. While at the degenerate points $k_x=\pm \pi/a$, $S^x_1 \approx -1$ and $S^x_2 \approx 1$ survive, noting the tilt angle $\theta \rightarrow 0$ by $J \gg D \gg |K_x| $ such that $A \gg B \rightarrow 0$ and $\alpha \rightarrow 0$. Since these spins are \textit{singularly} distributed in the BZ, we refer to them as the ``nodal" magnon spins.
Without the hidden DMI, we note that the two modes are degenerate in collinear antiferromagnet with magnon spins $S_x=\pm 1$.

Figure~\ref{1D}(b) and (c) illustrate the spin precession of the two modes. When non-degenerate, the precession of ${\bf S}_1$ and ${\bf S}_2$ in mode ``1" is out-of-phase while in mode ``2" is in phase since the local DMI couples the two antiferromagnetic modes without DMI to the antibonding and bonding states. When degenerate, the precession of the two spins in one magnetic unit cell decouples according to Hamiltonian (\ref{1D_H2}) since $C_{\pm}(k_x)=0$, as in Fig.~\ref{1D}(c).

 The global DMI cannot quench the magnon spins along the N\'eel vector but renders a spiral spin order with the period governed by the strength of the DMI. Generally, the period is relatively long when $D \ll J$, \textit{e.g.}, in the multiferroic material BiFeO$_3$ with a period length of $\sim 62$~nm~\cite{BFO1,BFO2,BFO3,BFO4,BFO5}. Such spiral spin orders are widely studied in two-dimensional materials~\cite{2D}.

The band degeneracy is protected by the crystal symmetries, which in this case is a twofold screw symmetry $\tilde{C}_{2y}$  that combines a $\pi$-rotation about the $\hat{\bf y}$-axis and a half-lattice translation along $\hat{\bf x}$. 
It acts on both the wave-vector and spin spaces:
$\Tilde{C}_{2y}|k_x\rangle\rightarrow e^{ik_xa/2}|-k_x\rangle$ and $\Tilde{C}_{2y}(S_x,S_y)\rightarrow(-S_x,S_y)$. We are interested in the mode degeneracy at the same wave vector, which corresponds to $k_x=0$ or $\pi/a$ since $\Tilde{C}_{2y}$ (equivalently) maps the states to themselves. When $k_x=0$, $\Tilde{C}_{2y}$ exchanges the sublattice spins $S_1 \rightarrow S_2$ such that $\Tilde{C}_{2y}\Psi_{1+}=-\Psi_{1+}$ and $\Tilde{C}_{2y}\Psi_{2+}=\Psi_{2+}$, i.e., $\Tilde{C}_{2y}$ maps the state to itself, providing no symmetry restriction. Nevertheless, when $k_x=\pm \pi/a$,   $\Tilde{C}_{2y}\Psi_{1+}=\Psi_{2+}$ connects the two states of frequencies $\omega_1$ and $\omega_2$, requesting that $\Psi_{1+}$ and $\Psi_{2+}$ are degenerate with the same frequencies.

Furthermore, magnon nodal spins are related to the symmetry protection. Without the local DMI, it is evident that the system possesses $PT$ symmetry in the collinear easy-axis antiferromagnet, where $P$ and $T$ represent the spatial inversion and time-reversal operations, respectively. The magnon has two degenerate eigenstates $|\pm \rangle$ with the spin $S_x=\pm 1$ along the easy-axis $\hat{\bf x}$.
Since the composite symmetry $PT$ inverts the spin of magnons, i.e., $PT |+ \rangle \rightarrow |- \rangle$, $|+\rangle$ and $|-\rangle$ are protected by this symmetry~\cite{CuTeO1}.
In addition, the system also satisfies the symmetry of $\Tilde{C}_{2y}$.
According to our understanding, when the local DMI is present, it breaks the symmetry $PT$ by tilting the ground-state magnetization, thus lifting generally the degeneracy of the magnon band, which causes the chiral modes $|\pm \rangle$ to be mixed and form linearly polarized modes $\Psi_{1+}=(|+ \rangle-|- \rangle)/\sqrt{2}$ and $\Psi_{2+}=(|+ \rangle+|- \rangle)/\sqrt{2}$, which carry vanishing spin on average. However, the symmetry $\Tilde{C}_{2y}$ is still preserved, protecting the energy degeneracy at the specific hot spots $k_x = \pm \pi/a$ in the Brillouin zone, which prevents $|+ \rangle$ and $|- \rangle$ from being mixed and allows the spin to survive.

In fact, the coupling that induces the bonding and anti-bonding states at the degenerate points may cause a similar configuration of magnon spins by, e.g., the proper magnetic field or local DMI.
Although magnetic fields along $\hat{\bf y}$ can produce similar configurations, the local DMI, as an intrinsic property of the material, allows non-collinear canted spin configurations to be stabilized without an external magnetic field. Such local DMI can be enhanced and manipulated by an electric field~\cite{DM3} that provides an electrical way of tuning magnon spin configuration.

\section{Uniaxial antiferromagnets}

\begin{figure}[htb!]	
\centering
\hspace*{-0.25cm}
\includegraphics[width=0.5\textwidth]{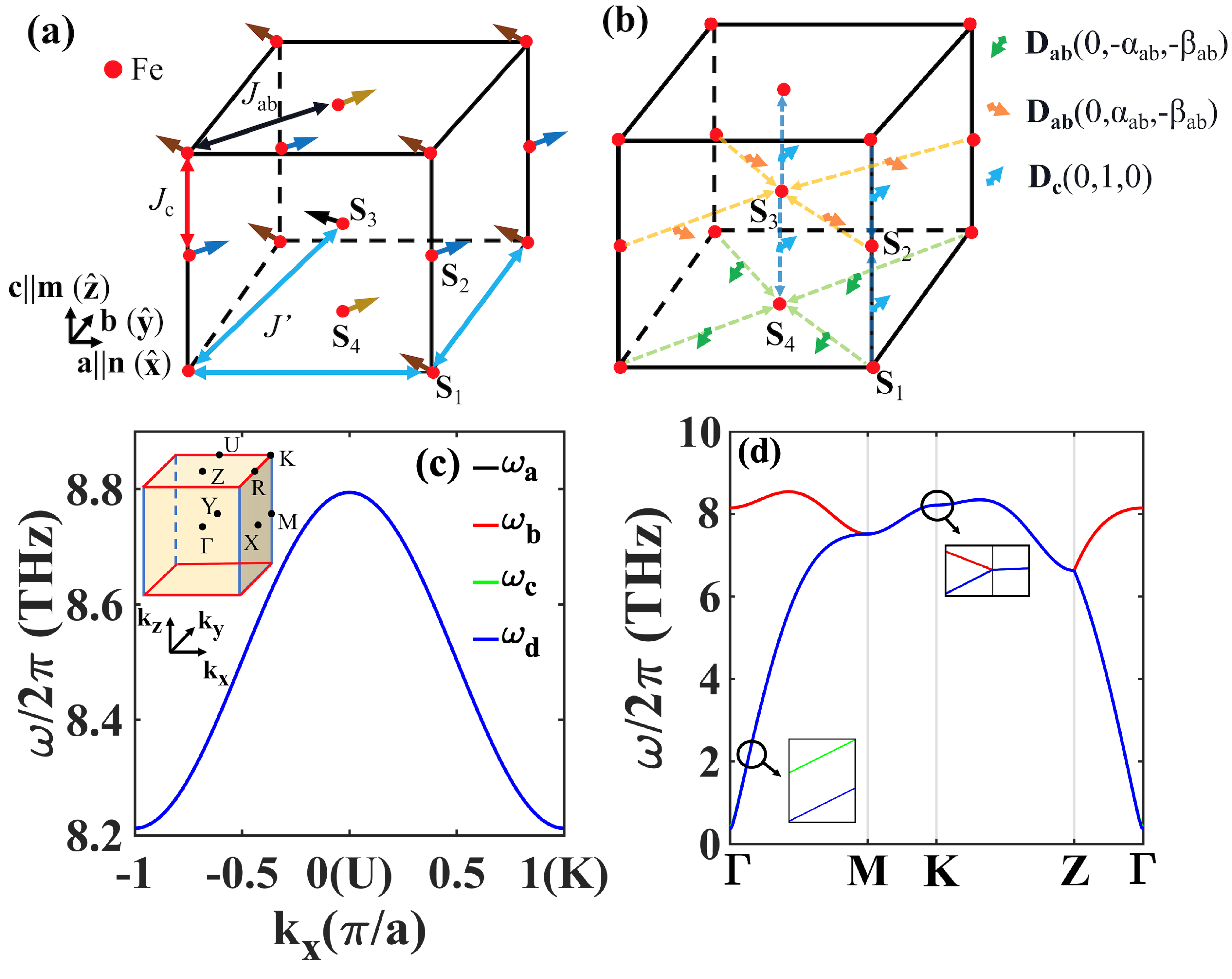}
\caption{(a) Ground-state configuration of $R$FeO$_3$ governed by the nearest-neighboring interaction in the $ab$-plane $J_{ab}$ and along the $c$-axis $J_c$, and next-nearest-neighboring interaction $J'$. (b) The DM vectors are represented by the solid arrows. The dashed arrows indicate the spin pairs. (c) The nodal line along $k_x$ (U-K) with $k_y=\pi/b$ and $k_z=\pi/c$. The inset in (c) indicates the high-symmetry points of the BZ. The red solid line denotes the position of the quadruple nodal line, enclosing the doubly degenerate nodal surface. (d) Band structure along the path of high-symmetry points $\Gamma$-M-K-Z-$\Gamma$, where K is visible as a quadruple degenerate point.}
\label{band}
\end{figure}

We now apply the general principle to realistic materials, e.g., rare-earth orthoferrite  $R$FeO$_3$~\cite{RFeO1,RFeO2,RFeO3,RFeO4,RFeO5,RFeO6,RFeO7,RFeO8}. There are four Fe$^{3+}$ sublattices in a magnetic unit cell, and by the local DMI their spins with amplitude $S=5/2$ are canted as in Fig.~\ref{band}(a),
which implies four magnon modes. 
The N\'eel vector ${\bf n}\parallel \hat{\bf x}$ is along the easy axis. We categorize the four different Fe$^{3+}$ spins in one magnetic unit cell as $\hat{{\bf S}}_{l=\{1,2,3,4\},i}$, in which ``$l$" and ``$i$" represent the Bravais sublattices and magnetic unit cell, respectively. As shown in Fig.~\ref{band}(a), the nearest-neighboring exchange interaction $J_c$ is along the $\hat{\bf c}$-axis and $J_{ab}$ lies in the $ab$-plane; and $J'$ represents the exchange constant along the next-nearest neighboring bonds. $K_a$ and $K_c$ represent the uniaxial anisotropies that stabilize the antiferromagnet along the $\hat{\bf a}(\hat{\bf x})$-axis and the ferromagnet along the $\hat{\bf c}(\hat{\bf z})$-axis.
Accordingly, the spin Hamiltonian of the antiferromagnetic $R$FeO$_3$ reads~\cite{RFeO2}
\begin{align}
\hat{H}&=\sum_{\langle i j\rangle} J_c\left(\hat{\bf S}_{1i} \cdot \hat{\bf S}_{2j}
+\hat{\bf S}_{3i} \cdot \hat{\bf S}_{4j}\right) \nonumber\\
&+\sum_{\langle i j\rangle} J_{a b}\left(\hat{\bf S}_{1i} \cdot \hat{\bf S}_{4j}+\hat{\bf S}_{2i} \cdot \hat{\bf S}_{3j}\right) \nonumber\\
&+\sum_{\langle\langle i j\rangle\rangle} J'\left[\left(\hat{\bf S}_{1i} \cdot \hat{\bf S}_{3j}+\hat{\bf S}_{2i} \cdot \hat{\bf S}_{4j}\right) +\sum_n^4 \left(\hat{\bf S}_{ni}\cdot\hat{\bf S}_{nj} \right)\right]\nonumber \\
&+\sum_{\langle i j\rangle} \left[\mathbf{D}_{ij} \cdot \left(\hat{\bf S}_{1i} \times \hat{\bf S}_{4j}\right) +\mathbf{D}'_{ij} \cdot \left(\hat{\bf S}_{2i} \times \hat{\bf S}_{3j}\right)\right. \nonumber\\
&\left.+D_c{\bf v}_{ij} \cdot \left(\hat{\bf S}_{1i} \times \hat{\bf S}_{2j}\right)+D_c{\bf v}_{ij} \cdot \left(\hat{\bf S}_{3i} \times \hat{\bf S}_{4j}\right)\right] \nonumber\\
&+\sum_{i}\sum_n^4 \left[K_a\left(\hat{S}_{ni}^x\right)^2 + K_c\left(\hat{S}_{ni}^z\right)^2\right],
\label{h1}
\end{align}
where $\langle\cdots \rangle$ and $\langle\langle\cdots \rangle\rangle$ represent the nearest and next-nearest neighboring sites.
Here, referring to the simplified model by Hahn \textit{et al}~\cite{RFeO2}, the DM vectors ${\bf D}_{ij}=D_{ab}{\bf u}_{ij}$ and ${\bf D}'_{ij}=D_{ab}{\bf u}'_{ij}$ couple the nearest-neighboring spins $\{ \hat{\bf S}_1,{ \hat{\bf S}}_4\}$ and $\{\hat{\bf S}_2,\hat{\bf S}_3\}$ in which the unit vectors ${\bf u}_{ij}=(0,-\alpha_{ab},-\beta_{ab})$ and ${\bf u}'_{ij}=(0,\alpha_{ab},-\beta_{ab})$ with  $\alpha_{ab}^2+\beta_{ab}^2=1$, and ${\bf v}_{ij}=(0,1,0)$ couples the spins $\{ \hat{\bf S}_1,{ \hat{\bf S}}_2\}$ and $\{ \hat{\bf S}_3,{ \hat{\bf S}}_4\}$. The direction of these DM vectors is represented by the solid arrows in Fig.~\ref{band}(b). For LaFeO$_3$, the direction cosine $\alpha_{ab}=0.67$ and  $\beta_{ab}=0.74$, and the magnitude of DMI $D_{ab}=0.107$~meV and $D_c=0.155$~meV. The other parameters $J_c=5.47$~meV, $J_{ab}=5.47$~meV, $J'=0.24$~meV, $K_a=-0.0124$~meV and $K_c=-0.0037$~meV~\cite{RFeO1}.

Without the local DMI (and a small easy-axis anisotropy along $\hat{\bf z}$), the bands would exhibit a fourfold degeneracy in the BZ surface and a twofold degeneracy in the BZ bulk, and the spin $S_x = \pm 1$ along the N\'eel vector would be independent of $\bf k$ due to the global $U(1)$ symmetry. The existence of hidden DMI  generally breaks these degeneracies~\cite{W.Lin,LaFeO1}. Still, one notable exception exists: the nodal line with fourfold degeneracy on the $k_z=\pi/c$ plane by the red solid lines in the inset of Fig.~\ref{band}(c) remains unaffected, as demonstrated in Fig.~\ref{band}(c) for the band structure along the U-K path.  The degeneracy on the BZ surface by the shaded regions in the inset of Fig.~\ref{band}(c) is twofold. Figure~\ref{band}(d) plots an overview of the band structure along the trajectory  $\Gamma$-M-K-Z-$\Gamma$ that links several high-symmetry points,  demonstrating the twofold degeneracy along the path M-K-Z (since these points lie on the surface of the BZ), and showing a fourfold degeneracy at corner K. It is expected that the magnon spins are generally quenched besides those with band degeneracies.

We substantiate these expectations by analyzing the magnon spins at the front and back surfaces of the BZ, where $k_y=\pm \pi/b$, $k_x\in(-\pi/a,\pi/a]$, and $k_z\in (-\pi/c,\pi/c]$.
The Hamiltonian $H_{k_y=\pm {\pi}/{b}}(k_x,k_z)=\begin{pmatrix}
{\cal H}(k_x,k_z) & 0 \\
0 & {\cal H}^{*}(k_x,k_z)
\end{pmatrix}$, which under the basis $(\hat{a}_{1,\bf k},\hat{a}_{2,\bf k},\hat{a}^{\dagger}_{1,\bf -k},\hat{a}^{\dagger}_{2,\bf -k})^T$
contains a block 
\begin{align}
{\cal H}(k_x,k_z)=\frac{S}{2}\begin{pmatrix}
\mathcal{A}(k_x) & \mathcal{C}(k_z) & \mathcal{B} & \mathcal{D}(k_z) \\
\mathcal{C}(k_z) & \mathcal{A}(k_x) & \mathcal{D}(k_z) & \mathcal{B} \\
\mathcal{B} & \mathcal{D}^{*}(k_z) & \mathcal{A}(k_x) & \mathcal{C}(k_z) \\
\mathcal{D}^{*}(k_z) & \mathcal{B} & \mathcal{C}(k_z) & \mathcal{A}(k_x)
\end{pmatrix},
\label{H_q}
\end{align}
where $\mathcal{A}(k_x)$, 
$\mathcal{B}$,
$\mathcal{C}(k_z)$, and $\mathcal{D}(k_z)$ are parameters listed in the SM~\cite{supplement}. 
The two positive eigenvalues of $\eta_0 {\cal H}(k_x,k_z)$ are
\begin{align}
    \hbar \omega_{a}&= ({S}/{2})\sqrt{(\mathcal{A}-\mathcal{C})^2-(\mathcal{B}-\mathcal{D})(\mathcal{B}-\mathcal{D}^{*})}, \nonumber\\ 
    \hbar \omega_{b}&= ({S}/{2})\sqrt{(\mathcal{A}+\mathcal{C})^2-(\mathcal{B}+\mathcal{D})(\mathcal{B}+\mathcal{D}^{*})},
\end{align}
and the other two modes with $\omega_c=\omega_a$ and $\omega_d=\omega_b$ are solved from $\eta_0 {\cal H}^{*}(k_x,k_z)$, suggesting that the energy bands are at least doubly degenerate on the BZ surface. We adopt again a hyperbolic parametrization by setting $\mathcal{A}+\mathcal{C}=l\cosh{\beta}$, $\mathcal{A}-\mathcal{C}=m\cosh{\alpha}$, $\mathcal{B}+\mathcal{D}=l\sinh{\beta}e^{i\gamma}$, and $\mathcal{B}-\mathcal{D}=m\sinh{\alpha}e^{i\gamma}$, with which $\hbar \omega_{a}=Sm/2$, $\hbar \omega_{b}=Sl/2$.

When the eigenvalues are not degenerate, i.e., $\omega_a \neq \omega_b$ when $k_z \neq \pm \pi/c$, the corresponding eigenvectors of $\eta{\cal H}(k_x,k_z)$ are
\begin{align}
    \Psi_{a+}&=\frac{\sqrt{2}}{2}\left(
\begin{array}{c}
\cosh{\frac{\alpha}{2}}e^{i\gamma} \\
-\cosh{\frac{\alpha}{2}}e^{i\gamma} \\
-\sinh{\frac{\alpha}{2}}\\
\sinh{\frac{\alpha}{2}}
\end{array}
\right),
\Psi_{b+}=\frac{\sqrt{2}}{2}\left(
\begin{array}{c}
-\cosh{\frac{\beta}{2}}e^{i\gamma} \\
-\cosh{\frac{\beta}{2}}e^{i\gamma} \\
\sinh{\frac{\beta}{2}}\\
\sinh{\frac{\beta}{2}}
\end{array}
\right), \nonumber\\ 
\Psi_{a-}&=\frac{\sqrt{2}}{2}\left(
\begin{array}{c}
\sinh{\frac{\alpha}{2}}e^{i\gamma} \\
-\sinh{\frac{\alpha}{2}}e^{i\gamma} \\
-\cosh{\frac{\alpha}{2}}\\
\cosh{\frac{\alpha}{2}}
\end{array}
\right), 
\Psi_{b-}=\frac{\sqrt{2}}{2}\left(
\begin{array}{c}
-\sinh{\frac{\beta}{2}}e^{i\gamma} \\
-\sinh{\frac{\beta}{2}}e^{i\gamma} \\
\cosh{\frac{\beta}{2}}\\
\cosh{\frac{\beta}{2}}
\end{array}
\right). 
\label{State3}
\end{align}
The eigenvectors of $\eta_0{\cal H}^{*}(k_x,k_z)$ are conjugate of $\Psi_{a\pm,b\pm}$.
According to Eq.~\eqref{eq_S}, when non-degenerate the spins along the N\'eel vector $\hat{\bf x}$ of modes ``$a$" and ``$b$" vanish: $S^x_a=S^x_b=0$.
Similar results are obtained for $S^x_c$ and $S^x_d$ by considering $\eta_0 {\cal H}^{*}(k_x,k_z)$.

Nevertheless, at $k_z=\pm \pi/c$, $\mathcal{C}=0$ and $\mathcal{D}=0$, indicating that the energy bands are quadruply degenerate with $\omega_a=\omega_b=\omega_c=\omega_d$.
In this situation, we consider applying a small magnetic field $H_x \hat{\bf x}$ along the N{\'e}el vector $\hat{\bf x}$-direction to lift the degeneracy and define the magnon spin without ambiguity. Incorporating the Zeeman coupling, 
the Hamiltonian $\mathcal{H}(k_x,k_z)$ with $k_z=\pm \pi /c$ under the basis $(\hat{a}_{1,{\bf k}},\hat{a}^{\dagger}_{1,-{\bf k}})^T$ and $(\hat{a}_{2,{\bf k}},\hat{a}^{\dagger}_{2,-{\bf k}})^T$ is decomposed into two diagonal blocks
\begin{align}
    \hat{H}_{a(b)}(k_x)&=\frac{S}{2}\begin{pmatrix}
        \mathcal{A}(k_x)\pm \mathcal{H}_{\rm Z} & \mathcal{B} \\
        \mathcal{B} & \mathcal{A}(k_x)\pm \mathcal{H}_{\rm Z} \\
        \end{pmatrix}, \nonumber
\end{align}
where $\mathcal{H}_{\rm Z}=g\mu_B \mu_0 H_x \cos\theta \cos\phi/S$ and $g$ is the Land{\'e} factor of electrons.
Due to the spin splitting, the frequencies of the two modes 
$\hbar \omega_{a(b)}=({S}/{2})\sqrt{(\mathcal{A}\pm \mathcal{H}_{\rm Z})^2-\mathcal{B}^2}$
are no longer degenerate. 
In terms of a hyperbolic parametrization $\{\alpha',\beta',m',l'\}$ with 
\begin{align}
    &\begin{cases} 
    \mathcal{A}+\mathcal{H}_{\rm Z}=m'\cosh{\alpha'} \\
    \mathcal{B}=m'\sinh{\alpha'}
 \end{cases},~~~~\begin{cases}
        \mathcal{A}-\mathcal{H}_{\rm Z}=l'\cosh{\beta'} \\
        \mathcal{B}=l'\sinh{\beta'} \nonumber
    \end{cases},
\end{align}
we find the corresponding eigenstates
\begin{align}
    \Psi_{a,+}&=\left(
    \begin{array}{c}
    -\cosh{\frac{\alpha'}{2}} \\
    \sinh{\frac{\alpha'}{2}}\\
    \end{array}
    \right),~~~~
     \Psi_{b,+}=\left(
    \begin{array}{c}
    -\cosh{\frac{\beta'}{2}} \\
    \sinh{\frac{\beta'}{2}}
    \end{array}
    \right),\nonumber\\
    \Psi_{a,-}&=\left(
    \begin{array}{c}
    -\sinh{\frac{\alpha'}{2}} \\
    \cosh{\frac{\alpha'}{2}}\\
    \end{array}
    \right),~~~~
    \Psi_{b,-}=\left(
    \begin{array}{c}
    -\sinh{\frac{\beta'}{2}} \\
    \cosh{\frac{\beta'}{2}}
    \end{array}
    \right). \nonumber
\end{align}
and the spins along the N\'eel vector
\begin{align}
   S^x_a&=\cos\theta \cos\phi \left(\cosh^2{\frac{\alpha'}{2}}+\sinh^2{\frac{\alpha'}{2}}\right), \nonumber \\
   S^x_b&=-\cos\theta\cos\phi\left(\cosh^2{\frac{\beta'}{2}}+\sinh^2{\frac{\beta'}{2}}\right).
\end{align}
When $H_x \rightarrow 0$, $(\alpha',\beta') \rightarrow \alpha$ and the modes ``$a$" and ``$b$" are degenerate.

\begin{figure}[htb!]	
\centering
\includegraphics[width=0.5\textwidth]{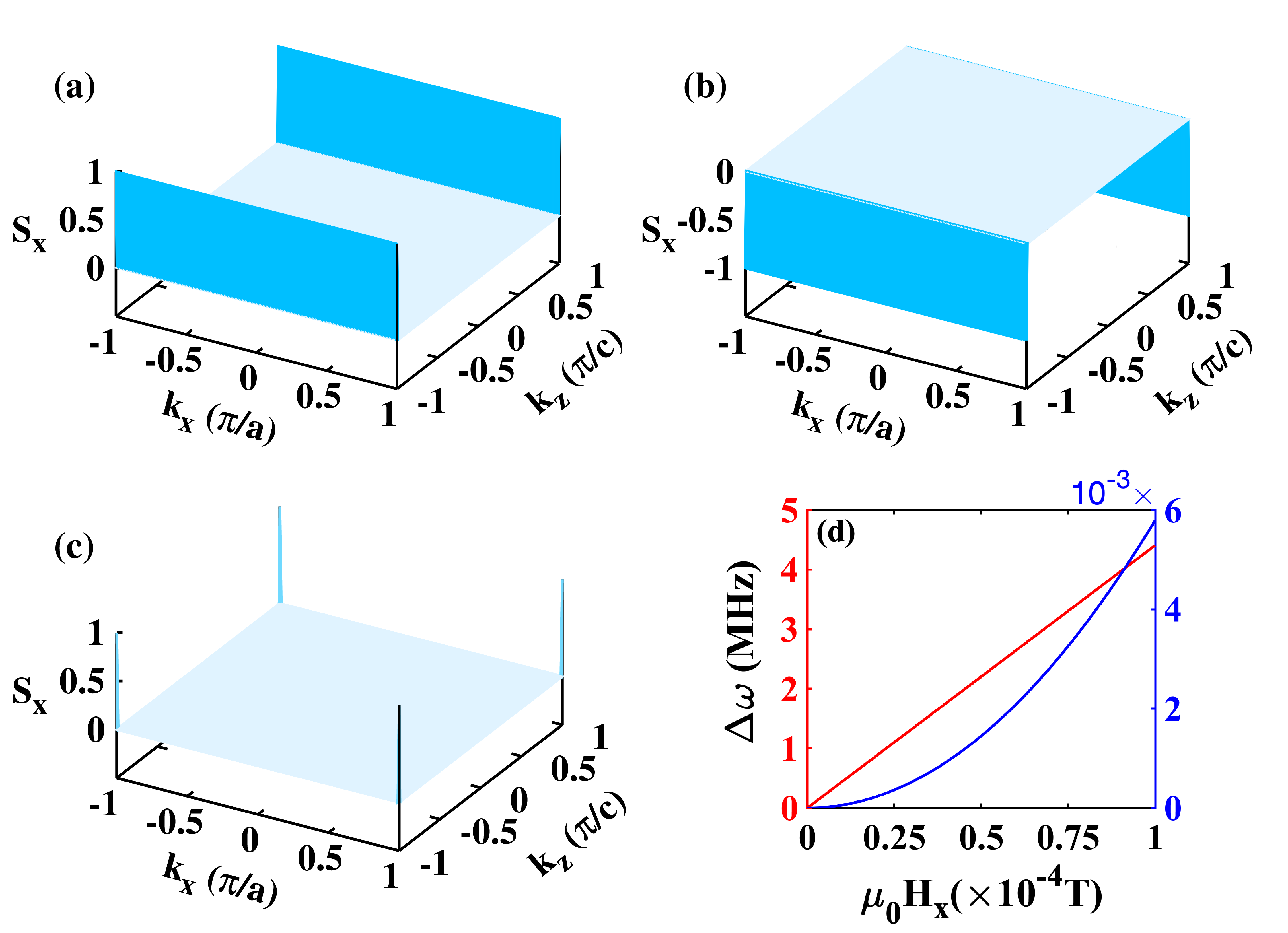}
\caption{Magnon spin $S_x$ of modes ``$a$" [(a)] and ``$d$" [(b)] in the plane of $k_y=\pi/b$. The spin $S_x\approx\pm 1$ only exists along the nodal line but is quenched everywhere else. (c) Corner spin $S_x(k_x,k_z)$ of mode ``$a$" in the plane of $k_y=0$. (d) Spin splitting by the magnetic field, governed by the Hellmann-Feynman theorem.}
\label{spin_H_0}
\end{figure}

In short, the spins of mode ``$a$" and ``$b$" read
\begin{align}
    S^x_a&=\left\{\begin{array}{cc}
    0,\hspace{4.9cm}\omega_a\ne \omega_b\\
    \cos\theta \cos\phi \left(\cosh^2{\frac{\alpha}{2}}+\sinh^2{\frac{\alpha}{2}}\right),~~~~\omega_a=\omega_b
    \end{array}\right., \nonumber \\
    S^x_b&=\left\{\begin{array}{cc}
    0,\hspace{4.9cm}\omega_a\ne \omega_b\\
    -\cos\theta\cos\phi\left(\cosh^2{\frac{\alpha}{2}}+\sinh^2{\frac{\alpha}{2}}\right),~~\omega_a=\omega_b
    \end{array}\right..
    \nonumber
\end{align}
Similar results are obtained for $S^x_c$ and $S^x_d$ by conjugation $\eta_0 {\cal H}^{*}(k_x,k_z)$.
At degeneracies $S^x_a=S^x_c \approx 1$ and $S^x_b=S^x_d \approx -1$ since the small tilt angles $\{\theta,\phi\}$ renders $\mathcal{A}\gg \mathcal{B} \rightarrow 0$ such that $\alpha \rightarrow 0$; besides, the spins are quenched. Such phenomena are well understood by the physics addressed in Fig.~\ref{1D}.
Figure~\ref{spin_H_0}(a) and (b) show the spin along the N\'eel vector $S_x$ for the high-energy mode ``$a$" and the low-energy mode ``$d$" in the $k_y=\pi/b$ plane,  which only exists along the nodal line.
Figure~\ref{spin_H_0}(c) shows the corner spin for the mode ``$a$" at the four corners intersecting with the nodal line in the $k_y=0$ plane.

The nodal and corner spins are protected again by the crystal symmetries including the spatial inversion $P$, the non-symmorphic twofold screw symmetry $\Tilde{C}_{2z}=\{C_{2z}|(0,0,1/2)\}$ in which $C_{2z}$ is a twofold rotation around the $\hat{\bf z}$-axis and $(0,0,1/2)$ is the translation of half unit cell along the $\hat{\bf z}$-axis, and the combination of two glide mirror operations with time-reversal symmetry ${\cal T}\Tilde{M}_{2x}={\cal T}\{C_{2x}|(1/2,1/2,0)\}$ and ${\cal T}\Tilde{M}_{2y}={\cal T}\{C_{2y}|(1/2,1/2,1/2)\}$~\cite{CuTeO1,Nodal1}. The symmetries ${\cal T}\Tilde{M}_{2x}$, ${\cal T}\Tilde{M}_{2y}$, and $\Tilde{R}_{2z}=P\Tilde{C}_{2z}$ lead, respectively, to the band double degeneracy of states at the planes of $k_x=\pm \pi/a$, $k_y=\pm \pi/b$, and $k_z=\pm \pi/c$. We refer the details of the symmetry analysis to Appendix~\ref{appendix_B}.

The magnon spins directly affect its spin splitting to the magnetic field, governed by the Hellmann-Feynman theorem~\cite{HF1,HF2}. Assuming a (weak) magnetic field applied along the N\'eel vector $\hat{\bf x}$-direction, according to the Hellmann-Feynman theorem
\begin{align}
    \mu_B S_x^{\xi}({\bf k})=(\hbar/\mu_0){\partial \omega_{\xi}({\bf k})}/{\partial H_x},
\end{align}
where $\mu_B$ is the Bohr magneton. It implies that for a finite $\partial \omega_{\xi}({\bf k})/\partial H_x$ the spin splitting is linear in $H_x$ with a slope $S_x$; when $S_x=0$, the spin splitting is at least quadratic in $H_x$. Figure~\ref{spin_H_0}(d) illustrates the spin splitting $\Delta \omega$ for mode ``$a$" in the $k_y=\pi/b$ plane, for which we average $\Delta \omega$ at the nodal line (the red curve) and the other positions (the blue curve). The spin splitting is linear in $H_x$ at the nodal line, indicating a finite $S_x$; besides, it is parabolic, indicating $S_x=0$.

\section{Spin transport enabled by magnetic field}

\begin{figure}[htb]	
\includegraphics[scale=0.38]{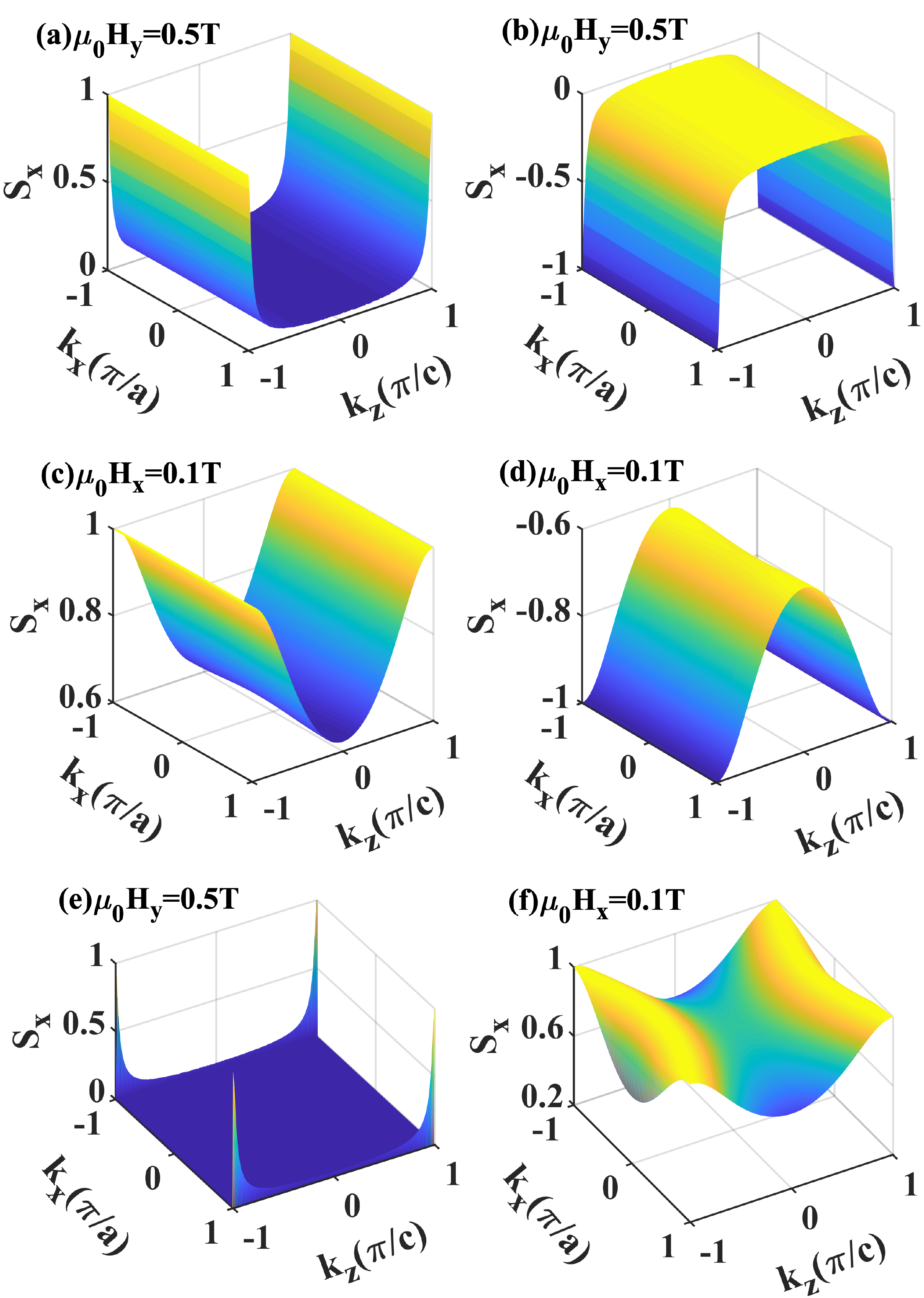}
\caption{\label{spin_H} Nodal magnon spins $S_x$ of modes ``$a$" [(a)] and ``$d$" [(b)] in the $k_y=\pi/b$ plane robust to the transverse magnetic field $H_y\hat{\bf y}$. In (c) and (d), nodal spins $S_x$ are broadened by the longitudinal field $H_x\hat{\bf x}$ along the N$\Acute{\rm e}$el vector. (e) and (f) show the corner spins $S_x$ of mode ``$a$" in the $k_y=0$ plane under $H_y\hat{\bf y}$ and  $H_x\hat{\bf x}$, respectively.} 
\end{figure}

The spin-quenching mechanism forbids the angular-momentum flow at zero magnetic fields in uniaxial antiferromagnets with hidden DMI. We further show that the nodal and corner magnon spins $S_x$ are broadened slightly by the transverse magnetic field along the $\hat{\bf y}$-direction but broadened strongly by the longitudinal field aligned with the N$\Acute{\rm e}$el vector $\hat{\bf x}$-direction, noting the effect of the small longitudinal magnetic field on the N{\'e}el vector rotation can be safely neglected, referring to the Supplemental Material~\cite{supplement} for details.
For the transverse field, the spins of the high-energy mode ``$a$" and low-energy mode ``$d$" at the $k_y=\pi/b$ plane are little changed, as shown in Fig.~\ref{spin_H}(a) and (b).
For the longitudinal field, on the other hand, the spin is broadened dramatically [Fig.~\ref{spin_H}(c) and (d)]. The corner spins, e.g., in the mode ``$a$", are similar: they are slightly affected by the transverse field while broadened strongly by the longitudinal field [Fig.~\ref{spin_H}(e) and (f)]. The longitudinal field along the N$\Acute{\rm e}$el vector also lifts the spin degeneracy, which remains under the transverse field.

The reappearance of magnon spins enabled by a magnetic field along the N$\Acute{\rm e}$el vector ${\bf n} \parallel\hat{\bf x}$ [Fig.~\ref{spin_H}(c,d,f)] is detectable in the non-local spin transport~\cite{YFeO,bart}. Here, we show the non-trivial role of the hidden DMI. A temperature gradient $\nabla_\beta T$ along the $\beta$-direction drives a longitudinal spin current $J^x_\beta=\sigma^x_\beta \nabla_\beta T$ for the spin along the N\'eel vector $\hat{\bf x}$, governed by the spin conductivity
\begin{align}
    \sigma^x_{\beta}=\sum_{\xi,{\bf k}} \frac{\tau}{Vk_B T^2}
    v^2_{\xi,\beta}({\bf k}) S^{x}_{\xi}({\bf k})  \frac{\hbar \omega_{\xi,{\bf k}} e^{{\hbar \omega_{\xi,{\bf k}}}/(k_B T)}}{(e^{{\hbar \omega_{\xi,{\bf k}}}/(k_B T)}-1)^2},
\end{align}
where $v_{\xi,\beta}({\bf k})$ is the group velocity of mode $\xi$ along the $\beta$-direction, $V$ is the crystal volume, and $\tau\sim 1$~ns is the scattering time estimated by the damping coefficient $\sim 5\times 10^{-6}$~\cite{YFeO}. To account for the temperature effect, we solve the thermal-averaged spin of Fe$^{3+}$ self-consistently via~\cite{J.M.D.Coey} 
\begin{align}
    S_{\rm eff}=\frac{1}{2}\left[(2S+1)\coth\frac{(2S+1)6JS_{\rm eff}}{2Sk_BT}-\coth\frac{6JS_{\rm eff}}{2Sk_BT}\right],
\end{align}
governed by the
nearest-neighboring exchange coupling $J=5.47$ (4.77)~meV  for LaFeO$_3$~\cite{RFeO1} (YFeO$_3$~\cite{RFeO2,RFeO3}).

\begin{figure}[htb]	
\includegraphics[scale=0.25]{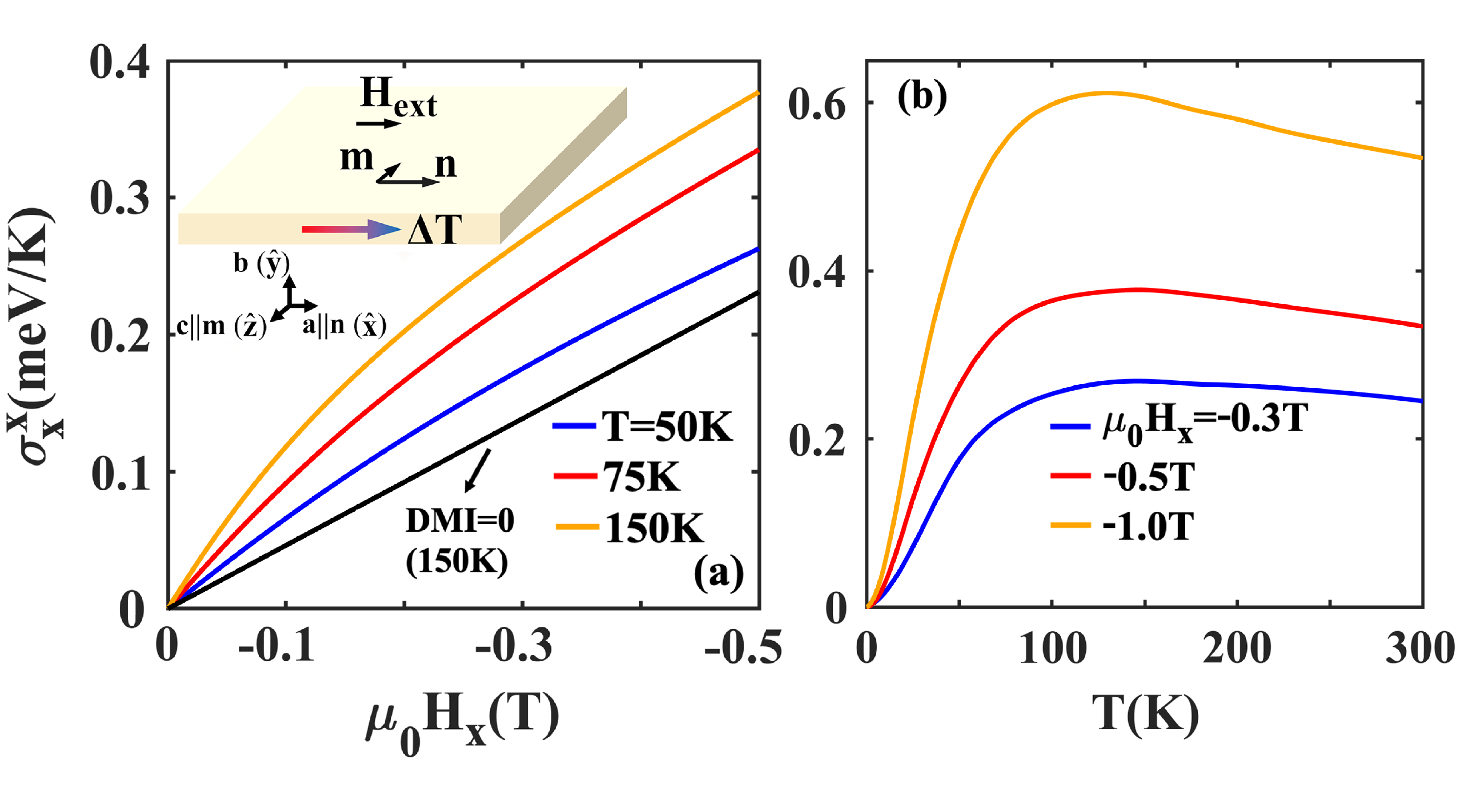}
\caption{\label{sigma} Longitudinal magnon spin conductivity when biased by a magnetic field along the N\'eel vector in the magnetic-field [(a)] and temperature [(b)] dependencies.} 
\end{figure}

As implied by the Hellmann-Feynman theorem in Fig.~\ref{spin_H_0}(d), the dependence of the thermal conductivity $\sigma^x$ on the magnetic field is not linear, as confirmed by Fig.~\ref{sigma}(a). 
This is in contrast to the uniaxial antiferromagnets without local DMI, in which $\sigma^x\propto H_x$ [the black curve in Fig.~\ref{sigma}(a)].
The hidden DMI renders the scaling law  $\sigma^x\propto H_x^n$, in which at $150$~K we find $n\approx 0.65$, which agrees excellently with the recent experiment $n\sim 0.6$~\cite{YFeO}. 
Figure~\ref{sigma}(b) shows a nontrivial peak of spin conductivity in the temperature dependence. With the increase of temperature, more magnon population $f_{\xi,{\bf k}}$ participate in the transport until all the high-energy modes are involved. On the other hand, the driven force $\partial f_{\xi,{\bf k}}/\partial T$ and the effective spin $S_{\rm eff}$ decreases at high temperatures, leading to the decrease of spin conductivity. The peak of spin conductivities in the temperature dependence is observed in several antiferromagnets~\cite{easy_axis4,easy_axis7,YFeO,easy_plane2}.

\section{Conclusion and discussion}

Local breaking inversion symmetry is a general phenomenon that induces local or hidden DMI in a large class of uniaxial antiferromagnets, which we predict causes anomalous phenomena in magnon spin configuration and transport. Even when weak, the magnon spins along the  N\'eel vector are generally quenched in these canted antiferromagnets but survive at high-symmetric hot points as nodal or corner spins. These spin distributions, when broadened by the magnetic field along the N\'eel vector, are responsible for bulk spin transport with unique signatures in the magnetic field and temperature dependencies that can be detected. Our work highlights the role of hidden interaction in magnetism that may inspire other investigations.

\begin{acknowledgments}
This work is financially supported by the National Key Research and Development Program of China under Grant No.~2023YFA1406600, the National Natural Science Foundation of China under Grants No.~12374109 and No.~12074065, the open research fund of Key Laboratory of Quantum Materials and Devices of Ministry of Education, as well as the startup grant of Huazhong University of Science and Technology. 
\end{acknowledgments}

\begin{appendix}
    
\section{Magnon spins in a general model}
\label{appendix_A}

The Heisenberg Hamiltonian in the spin-lattice model can be generally expressed as
\begin{align}
    \hat{H}=\frac{1}{2}\sum_{l,l'}^{\cal N}\sum_{i,j}^{N}\sum_{\alpha,\beta}{\cal J}_{ij}^{\alpha \beta}(l,l')\hat{S}_{li}^{\alpha}\hat{S}_{l'j}^{\beta},
    \label{Hamiltonian_general}
\end{align}
in the summation over the unit cells $\{i,j\}$ with $N$ being their total number in the lattice, the spin sublattices $\{l,l'\}$ with ${\cal N}$ being their total number in one unit cell, and the spin (or spatial) indices $\{\alpha,\beta\}=\{x,y,z\}$. In the Hamiltonian \eqref{Hamiltonian_general}, ${\cal J}$ represents the interaction between spins $\hat{\bf S}$.

The Holstein-Primakoff (HP) transformation is based on the ground-state magnetic configurations, in which the direction of the spins are locally set to be along the $\hat{\bf z}'$-axis, which in the non-collinear magnets invokes the local frames. The ${\cal N}$ sublattices with spins along different directions in one magnetic unit cell then imply ${\cal N}$ local frames. 
Accordingly, for the $l$-th magnetic ions in a unit cell, we define a local $(x'_l,y'_l,z'_l)$-frame, where $\hat{\bf z}'_l$ is along the magnetization direction of the $l$-th magnetic ions. We then invoke a set of rotation matrices that transform from the local $\{\hat{\bf x}',\hat{\bf y}',\hat{\bf z}'\}$ to the lab $\{\hat{\bf x} \parallel \hat{\bf a}, \hat{\bf y} \parallel \hat{\bf b},\hat{\bf z} \parallel \hat{\bf c}\}$ frame,  where $\{\hat{\bf a},\hat{\bf b},\hat{\bf c}\}$ refers to three crystal main axes, according to 
\begin{equation}
\left(\begin{array}{c}\hat{S}_{li}^x \\
 \hat{S}_{li}^y \\
\hat{S}_{li}^z\end{array}\right)
=\left({\bf x}'_l,{\bf y}'_l,{\bf z}'_l\right) \left(\begin{array}{c}\hat{S}_{li}^{x'} \\
\hat{S}_{li}^{y'} \\ \hat{S}_{li}^{z'}\end{array}\right).
\label{S-S'}
\end{equation}
We introduce the bosonic operators $\hat{a}_{li}$ in the local $\{\hat{\bf x}'_l,\hat{\bf y}'_l,\hat{\bf z}'_l\}$-frame via the HP transformation
\begin{equation}
\begin{aligned}
\hat{S}_{li}^{x'}&=\frac{\sqrt{2 S}}{2}\left(\hat{a}_{li}+\hat{a}_{li}^{\dagger}\right), \\
 \hat{S}_{li}^{y'}&=\frac{\sqrt{2 S}}{2 i}\left(\hat{a}_{li}-\hat{a}_{li}^{\dagger}\right), \\
 \hat{S}_{li}^{z'}&=S-\hat{a}_{li}^{\dagger} \hat{a}_{li},
\end{aligned}
 \end{equation}
which when transformed to the lab frame by Eq.~(\ref{S-S'}) yields~\cite{S.Toth,S.Petit}
\begin{equation} \hat{S}_{li}^{\alpha=\{x,y,z\}}=\frac{\sqrt{2S}}{2} \xi_l^{\alpha *} \hat{a}_{li}+\frac{\sqrt{2 S}}{2} \xi_l^\alpha \hat{a}_{li}^{\dagger}+\eta_l^\alpha\left(S-\hat{a}_{li}^{\dagger} \hat{a}_{li}\right), 
\label{h2}
\end{equation}
where ${\bm \xi}_l={\bf x}'_l+i {\bf y}'_l$ and ${\bm \eta}_l={\bf z}'_l$.
After the Fourier transformation over the unit cells $\{i,j\}$, we obtain the Hamiltonian
 \begin{align}
   \hat{H}=\frac{1}{2}\sum_{\bf k} \hat{X}_{\bf k}^{\dagger}H_{\bf k}\hat{X}_{\bf k},
   \label{Hamiltonian}
 \end{align}
in which $\hat{X}_{\bf k}=\left(\hat{a}_{l,{\bf k}}, \\ \hat{a}^{\dagger}_{l,{\bf -k}} \right)^T$ with $\{l=1,...,{\cal N}\}$ and the $2{\cal N} \times 2{\cal N}$ Hamiltonian matrix 
\begin{equation}
    H_{\bf k}=\begin{pmatrix}
        h_{\bf k}|_{{\cal N}\times {\cal N}} & \Delta_{\bf k}|_{{\cal N}\times {\cal N}}\\
        \Delta^{*}_{\bf -k}|_{{\cal N}\times {\cal N}} & h^{*}_{\bf -k}|_{{\cal N}\times {\cal N}}
    \end{pmatrix}.
    \label{H_matrix}
\end{equation}

We then perform the Bogoliubov transformation for the Hamiltonian matrix \eqref{H_matrix}. To this end, we find a $2{\cal N} \times 2{\cal N}$ transformation matrix $T_{\bf k}$ such that
\begin{equation}
    \hat{X}_{\bf k}=T_{\bf k}\hat{X}_{\bf k}',
    \label{transformation}
\end{equation}
where $\hat{X}_{\bf k}^{\prime}=\left(\hat{b}_{\xi, {\bf k}}, \\ \hat{b}^{\dagger}_{\xi, {\bf -k}}\right)^T$ are the magnon operators of modes $\xi=\{1,...,{\cal N}\}$, rendering the Hamiltonian \eqref{Hamiltonian}  
\begin{equation}
    \hat{H}=\frac{1}{2}\sum_{\bf k}\hat{X}_{\bf k}'^{\dagger}T_{\bf k}^{\dagger}H_{\bf k}T_{\bf k}\hat{X}_{\bf k}',
\end{equation}
in which the matrix 
\begin{align}
    T_{\bf k}^{\dagger}H_{\bf k}T_{\bf k}={\rm diag}(\omega_{1,\bf k},...,\omega_{{\cal N},\bf k},\omega_{1,\bf -k},...,\omega_{{\cal N},\bf -k})
    \label{diagonal}
\end{align}
is diagonal in terms of the magnon dispersion $\omega_{\xi,{\bf k}}$. 
The operators before and after the transformation \eqref{transformation} are required to obey the boson commutation relation such that 
\begin{equation}\label{h6}
    \left[\hat{X}_{\bf k},\hat{X}_{\bf k}^{\dagger}\right]=\left[\hat{X}_{\bf k}',\hat{X}_{\bf k}'^{\dagger}\right]=\eta,
\end{equation}
where the metric matrix 
$\eta=\begin{pmatrix} 
I_{{\cal N} \times {\cal N}} & 0 \\
0 & -I_{{\cal N} \times {\cal N}}
\end{pmatrix}$.
Substitution of Eq.~(\ref{transformation}) into (\ref{h6}) leads to 
\begin{align}
\left[\hat{X}_{\bf k},\hat{X}_{\bf k}^{\dagger}\right]&=\hat{X}_{\bf k} \hat{X}_{\bf k}^{\dagger}-\left(\hat{X}_{\bf k}^{*} \hat{X}_{\bf k}^{T}\right)^{T} \nonumber \\
    &=T_{\bf k}\hat{X}_{\bf k}'\hat{X}_{\bf k}'^{\dagger}T_{\bf k}^{\dagger}
    -\left((T_{\bf k}\hat{X}_{\bf k}')^{*}(T_{\bf k}\hat{X}_{\bf k}')^T\right)^T \nonumber\\
    &=T_{\bf k}\hat{X}_{\bf k}'\hat{X}_{\bf k}'^{\dagger}T_{\bf k}^{\dagger}
    -T_{\bf k}(\hat{X}^{'*}_{\bf k}\hat{X}^{'T}_{\bf k})^T T_{\bf k}^{\dagger} \nonumber\\
    &=T_{\bf k} \left[\hat{X}_{\bf k}',\hat{X}_{\bf k}'^{\dagger}\right] T_{\bf k}^{\dagger}
    =\eta,
\end{align}
from which we find the relation 
\begin{equation}
    T_{\bf k}\eta T_{\bf k}^{\dagger}=\eta~~\mathrm{or}~~T_{\bf k}^{\dagger}\eta T_{\bf k}=\eta,
    \label{h7}
\end{equation}
in which the second relation can be derived from the first one according to $(T_{\bf k} \eta)T_{\bf k}^{\dagger} \eta=I$, i.e., $T_{\bf k}^{\dagger} \eta = \eta T_{\bf k}^{-1}$.

On the other hand, based on the Heisenberg equation of motion, we obtain the Schr\"odinger-like equation for operators $\hat{X}_{\bf k}$:
\begin{equation}
  i \partial_t \hat{X}_{\bf k}=\left[\hat{X}_{\bf k},\hat{H}\right]=(\eta H_{\bf k})\hat{X}_{\bf k},
  \label{h8}
\end{equation}
in which $\eta H_{\bf k}$ plays the role of an ``Hermitian Hamiltonian" matrix. 
Similarly, for operators $\hat{X}'_{\bf k}$, we find 
\begin{equation}
\label{h9}
  \begin{aligned}
  i \partial_t \hat{X}_{\bf k}' &=T_{\bf k}^{-1}i \partial_t \hat{X}_{\bf k}=T_{\bf k}^{-1} (\eta H_{\bf k}) \hat{X}_{\bf k} \\
  &=T_{\bf k}^{-1} (\eta H_{\bf k}) T_{\bf k}\hat{X}_{\bf k}'.
  \end{aligned}
\end{equation}
Since $\hat{X}_{\bf k}'$ is the magnon operator, the matrix $T_{\bf k}^{-1} (\eta H_{\bf k}) T_{\bf k}$ should be diagonal.
With the first term of Eq.~(\ref{h7}), $\eta T_{\bf k}^{\dagger}=T_{\bf k}^{-1} \eta$ such that $\eta T_{\bf k}^{\dagger} H_{\bf k}=T_{\bf k}^{-1} \eta H_{\bf k}$ and 
$\eta T_{\bf k}^{\dagger} H_{\bf k} T_{\bf k}=T_{\bf k}^{-1} \eta H_{\bf k} T_{\bf k}$. Then from Eq.~\eqref{diagonal}, we arrive at 
\begin{align}
T_{\bf k}^{-1} (\eta H_{\bf k}) T_{\bf k}={\rm diag}(\omega_{1,\bf k},...,\omega_{{\cal N},\bf k},-\omega_{1,\bf -k},...,-\omega_{{\cal N},\bf -k}),
\nonumber
\end{align}
in which the positive and negative eigenvalues appear in pairs. Accordingly, the eigenvectors $u_i$ of $\eta H_{\bf k}$ can be arranged to express the transformation matrix $T_{\bf k}$. With the second relation in Eq.~(\ref{h7}), these eigenvectors are para-perpendicular to each other with $u_i^{\dagger} \eta u_j=0$ and para-normalized with $u_i^{\dagger} \eta u_i=\pm 1$, in which ``$+$" (``$-$") indicates the index $i=\{1,..,{\cal N}\}$ ($i=\{{\cal N}+1,...,2{\cal N}\}$)~\cite{J.H.P.Colpa}.

We then derive the spin carried by each magnon mode \eqref{h2}. The $\alpha$-component of the total-spin operator
\begin{align}
\hat{S}^{\alpha}=\sum_{l=1}^{\cal N} \sum_i \hat{S}_{li}^{\alpha}
 &=-\sum_{l=1}^{\cal N} \sum_{\bf k}\eta_l^{\alpha} \hat{a}^{\dagger}_{l,{\bf k}} \hat{a}_{l,{\bf k}} \nonumber\\
 &+{\rm zeroth ~ and ~ first ~ order ~ terms ~ of ~}\{\hat{a},\hat{a}^{\dagger}\},
 \label{total_spin}
 \end{align} 
in which the zeroth and first orders of the bosonic operators have no contribution to the magnon spin. Substituting the magnon operators 
\begin{align}
    \hat{a}_{l,{\bf k}}&=\sum_{\xi=1}^{\cal N}\left(T_{l,\xi}({\bf k})\hat{b}_{\xi,{\bf k}} +T_{l,\xi+{\cal N}}({\bf k})\hat{b}^{\dagger}_{\xi,{\bf -k}}\right),\nonumber \\
    \hat{a}^{\dagger}_{l,{\bf k}} 
    &=\sum_{\xi=1}^{\cal N}\left(T^{\ast}_{l,\xi}({\bf k})\hat{b}^{\dagger}_{\xi,{\bf k}}+T^{\ast}_{l,\xi+{\cal N}}({\bf k})\hat{b}_{\xi,{\bf -k}}\right), 
\end{align}
into \eqref{total_spin}, we obtain  
\begin{align}
    \hat{S}^{\alpha}
    &=-\sum_{{\bf k}}\sum_{l=1}^{\cal N}\sum_{\xi=1}^{\cal N} \eta_l^{\alpha}\left(|T_{l,\xi}({\bf k})|^2+
    |T_{l,\xi+{\cal N}}({\bf -k})|^2\right)\hat{b}^{\dagger}_{\xi,{\bf k}}\hat{b}_{\xi,{\bf k}} \nonumber \\
    &+ {\rm off~diagonal ~ terms}.
    \label{total_spin_2}
\end{align}
Thus, the spin of magnon $|\xi,{\bf k} \rangle=b^{\dagger}_{\xi, \bf k} |0 \rangle$ of the $\xi$-mode with wave vector ${\bf k}$ reads~\cite{N.Okuma}
\begin{align}
   S^{\alpha}_{\xi,\bf k}&=\langle \xi,{\bf k} |\hat{S}^{\alpha}|\xi,{\bf k} \rangle
   -\langle 0 |\hat{S}^{\alpha}|0 \rangle \nonumber \\
   &=-\sum_l^{\cal N} \eta_l^{\alpha}
    \left(|T_{l,\xi}({\bf k})|^2+|T_{l,\xi+{\cal N}}({\bf -k})|^2 \right),
    \label{eq_S}
\end{align}
in which the ``off-diagonal terms" in \eqref{total_spin_2} have no net contribution,
where $|0 \rangle$ is the Fock vacuum of the magnon operator.

\section{Symmetry analysis for $R$FeO$_3$ mode degeneracies}
\label{appendix_B}

As we address in this part, the band degeneracies and the nodal magnon spins are protected by the crystal symmetries.

In the case of $R$FeO$_3$, the following symmetries are of particular importance for the magnon bands: the spatial inversion symmetry $P$, the non-symmorphic twofold screw symmetry $\Tilde{C}_{2z}=\{C_{2z}|(0,0,1/2)\}$, in which $C_{2z}$ is a twofold rotation around the $\hat{\bf z}$-axis and $(0,0,1/2)$ is the translation of half unit cell along the $\hat{\bf z}$-direction, and the combination of two glide mirror operations with time-reversal symmetry, i.e., ${\cal T}\Tilde{M}_{2x}={\cal T}\{C_{2x}|(1/2,1/2,0)\}$ and ${\cal T}\Tilde{M}_{2y}={\cal T}\{C_{2y}|(1/2,1/2,1/2)\}$. 
It should be emphasized that the system does not exhibit strict time-reversal symmetry in the presence of magnetic orders. However, the combination of ${\cal T}$ with certain spatial transformations can give rise to an \textit{effective} time-reversal symmetry, and such operation is antiunitary, i.e., the combined symmetry includes a complex conjugation operation that is brought about by the time-reversal ${\cal T}$, rendering ${\bf k} \rightarrow {\bf -k}$ and complex conjugation of matrix elements.

Each symmetry operation acts on both the real and spin spaces, which maps the states in the ${\bf k}$-space. Generally, a state at ${\bf k}$ is mapped to the other state at ${\bf k'}$ and they have the same energy.
When analyzing the band degeneracies, we only need to focus on the states that are mapped to the same (or equivalent) ${\bf k}$, which implies the same states or the band degeneracies \textit{when they are different states}. In such cases, a detailed analysis based on the specific form of the states is required. We find the symmetries ${\cal T}\Tilde{M}_{2x}$, ${\cal T}\Tilde{M}_{2y}$, and $\Tilde{R}_{2z}=P\Tilde{C}_{2z}$  lead, respectively, to the band double degeneracies in the $k_x=\pm \pi/a$, $k_y=\pm \pi/b$, and $k_z=\pm \pi/c$ planes \cite{CuTeO1,Nodal1}, as explained in the following.

Firstly, the operations ${\cal T}\Tilde{M}_{2y}$ and ${\cal T}\Tilde{M}_{2x}$  acting on the real and spin spaces lead to 
\begin{align}
    {\cal T}\Tilde{M}_{2y}:~&(x,y,z,t) \rightarrow \left(-x+\frac{a}{2}, y+\frac{b}{2}, -z+\frac{c}{2},-t\right), \nonumber \\
    &(S_x,S_y,S_z) \rightarrow (S_x, -S_y, S_z),\nonumber\\ 
    {\cal T}\Tilde{M}_{2x}:~&(x,y,z,t) \rightarrow \left(x+\frac{a}{2}, -y+\frac{b}{2}, -z,-t\right),\nonumber\\
    &(S_x,S_y,S_z) \rightarrow (-S_x, S_y, S_z).
\end{align}
For the magnon states $|k_x,k_y,k_z\rangle$ under the operation of ${\cal T}\Tilde{M}_{2y}$, we find 
\begin{align}
    {\cal T}\Tilde{M}_{2y}|k_x,k_y,k_z\rangle&={\cal T}e^{ik_xa/2}e^{ik_yb/2}e^{ik_zc/2}|-k_x,k_y,-k_z\rangle \nonumber\\
    &=e^{ik_xa/2}e^{ik_yb/2}e^{ik_zc/2}|k_x,-k_y,k_z\rangle^*, \nonumber\\
    (S_x,S_y,S_z)& \rightarrow (S_x,-S_y,S_z).
    \label{TM_2y}
\end{align}
In the first expression, the operation is equivalent to a translation operation in the real space, \textit{e.g.}, $e^{ik_xa/2}$ represents a translation $a/2$ along the $\hat{\bf x}$-axis.
Performing the operation twice leads to $({\cal T}\Tilde{M}_{2y})^2|k_x,k_y,k_z\rangle=e^{ik_yb}|k_x,k_y,k_z\rangle$ and $(S_x,S_y,S_z)\rightarrow(S_x,S_y,S_z)$ since
\begin{align}
    (x,y,z,t) &\xrightarrow{{\cal T}\Tilde{M}_{2y}} \left(-x+\frac{a}{2}, y+\frac{b}{2}, -z+\frac{c}{2},-t\right) \nonumber\\
    &\xrightarrow{{\cal T}\Tilde{M}_{2y}} \left(x, y+b, z,t\right), \nonumber \\
    (S_x,S_y,S_z) &\xrightarrow{{\cal T}\Tilde{M}_{2y}} (S_x, -S_y, S_z) \xrightarrow{{\cal T}\Tilde{M}_{2y}} (S_x,S_y,S_z),
\end{align}
or by a direct derivation in the {\bf k}-space,
\begin{align}
    &e^{ik_xa/2}e^{ik_yb/2}e^{ik_zc/2}{\cal T}\Tilde{M}_{2y}|k_x,-k_y,k_z\rangle^* \nonumber\\
    &=e^{ik_yb}{\cal T}|-k_x,-k_y,-k_z\rangle^* \nonumber\\
    &=e^{ik_yb}|k_x,k_y,k_z\rangle.
\end{align}
When $k_y=\pm \pi/b$, $e^{ik_yb}=-1$, implying a Kramers degeneracy of the states on this Brillouin zone surface \cite{Nodal1}. This is because when $k_y=\pm \pi/b$, the transformation ${\cal T}\Tilde{M}_{2y}$ in the ${\bf k}$-space maps the wave vector back to itself, but in the spin space, it swaps the spins on the sublattices (${\bf S}_1\rightarrow {\bf S}_3$ and ${\bf S}_2 \rightarrow {\bf S}_4$), i.e., the operation maps two different states. 
We can explicitly show the operation maps two different states: according to Eq.~(\ref{State3}), ${\cal T}\Tilde{M}_{2y}\Psi_{a+}=\Psi_{a+}^*$. Here, $\Psi_{a+}$ and $\Psi_{a+}^{*}$ are two different states with the same eigenenergy $ \hbar \omega_a=Sm/2$, which belong to the eigenstates of $\eta_0{\cal H}$ and $\eta_0{\cal H}^*$, respectively. The analysis for $\Psi_{b+}$ are similar. We conclude that the ${\cal T}\Tilde{M}_{2y}$ symmetry protects the twofold degeneracy of the four magnon bands.

Similarly, for the operation ${\cal T}\Tilde{M}_{2x}$,
\begin{align}
    {\cal T}\Tilde{M}_{2x}|k_x,k_y,k_z\rangle&={\cal T}e^{ik_xa/2}e^{ik_yb/2}|k_x,-k_y,-k_z\rangle \nonumber\\
    &=e^{ik_xa/2}e^{ik_yb/2}|-k_x,k_y,k_z\rangle^*, \nonumber\\
    (S_x,S_y,S_z)& \rightarrow (-S_x,S_y,S_z).
    \label{TM_2x}
\end{align}
Applying the operation twice gives rise to $({\cal T}\Tilde{M}_{2x})^2|k_x,k_y,k_z\rangle=e^{ik_xa}|k_x,k_y,k_z\rangle$ and $(S_x,S_y,S_z)\rightarrow(S_x,S_y,S_z)$.
Again, there is a Kramers degeneracy of the states when $k_x=\pm \pi/a$ because $e^{ik_xa}=-1$. Similar to ${\cal T}\Tilde{M}_{2y}$, for the states at $k_x=\pm \pi/a$, the transformation ${\cal T}\Tilde{M}_{2x}$ in the ${\bf k}$-space maps the wave vector back to itself, but in the spin space, the spins are exchanged on the sublattices (${\bf S}_1 \rightarrow {\bf S}_4$ and ${\bf S}_2 \rightarrow {\bf S}_3$). 
To be specific, since the Hamiltonian at the $k_x=\pm \pi/a$ plane is the same as that at the $k_y=\pm \pi/b$ plane with $\cos{(k_xa)}$ replaced by $\cos{(k_yb)}$, according to Eq.~(\ref{State3}), ${\cal T}\Tilde{M}_{2x}\Psi_{a+}=-\Psi_{a+}^{*}$. Here, $\Psi_{a+}$ and $\Psi_{a+}^{*}$ are different states with the same eigenenergy $\hbar\omega_{a}=Sm/2$, which are the eigenstates of $\eta {\cal H}$ and $\eta {\cal H}^*$, respectively. The transformation for $\Psi_{b+}$ is similar, thus ${\cal T}\Tilde{M}_{2x}$ protects the twofold degeneracy of the four magnon bands at the $k_x=\pm \pi/a$ plane.

We then consider the operation $\Tilde{R}_{2z}=P\Tilde{C}_{2z}$. We find it leads to the twofold degeneracy in the $k_z=\pm \pi/c$ plane. When $\Tilde{R}_{2z}$ acts on the real and spin spaces simultaneously, 
\begin{align}
    \Tilde{R}_{2z}:~&(x,y,z) \rightarrow (x,y,-z-\frac{c}{2}), \nonumber\\ 
    &(S_x,S_y,S_z) \rightarrow (-S_x, -S_y, S_z).
\end{align}
Accordingly, in the ${\bf k}$-space, we find
\begin{align}
\Tilde{R}_{2z}|k_x,k_y,k_z\rangle &=e^{-ik_zc/2}|k_x,k_y,-k_z\rangle, \nonumber\\
(S_x,S_y,S_z)& \rightarrow (-S_x, -S_y, S_z).
    \label{Rz}
\end{align}
The role of $\Tilde{R}_{2z}$ is analogous to that of $\Tilde{C}_{2y}$ for the one-dimensional spin chain discussed in the main text.
At $k_z=\pm \pi/c$, the transformation $\Tilde{R}_{2z}$ in the ${\bf k}$-space maps the wave vector back to itself, but in the spin space, it involves the exchange of spins on the sublattices (${\bf S}_1 \rightarrow {\bf S}_2$ and ${\bf S}_3 \rightarrow {\bf S}_4$). Similar to the above arguments, this leads to the double degeneracy of magnon bands at the plane of $k_z=\pm \pi/c$, i.e., $\omega_a=\omega_b$ and $\omega_c=\omega_d$.

At the edge of the Brillouin zone, we now show that the combination of $\Tilde{R}_{2z}$ with ${\cal T}\Tilde{M}_{2y}$ or ${\cal T}\Tilde{M}_{2x}$ protects a nodal line with fourfold degeneracy.
As indicated in the previous analysis, at the surfaces with $k_y=\pm \pi/b$ ($k_x=\pm \pi/a$) of the Brilloin zone, under the operations ${\cal T}\Tilde{M}_{2y}$ (${\cal T}\Tilde{M}_{2x}$), the basis of the states $(\hat{a}_{1,{\bf k}}, \hat{a}^{\dagger}_{1,{\bf -k}},\hat{a}_{2,{\bf k}},\hat{a}^{\dagger}_{2,{\bf -k}})^T \rightarrow (\hat{a}_{3,{\bf k}},\hat{a}^{\dagger}_{3,{\bf -k}}, \hat{a}_{4,{\bf k}},\hat{a}^{\dagger}_{4,{\bf -k}})^T$. Since there exists the double degeneracy between the two subspaces ($\omega_a=\omega_c$ and $\omega_b=\omega_d$),  this indicates that modes $\{\hat{a}_1,\hat{a}_2\}$ decouples with $\{\hat{a}_3,\hat{a}_4\}$.
On the other hand, $\Tilde{R}_{2z}$ transforms the basis of states $(\hat{a}_{1,{\bf k}},\hat{a}^{\dagger}_{1,{\bf -k}}, \hat{a}_{3,{\bf k}},\hat{a}^{\dagger}_{3,{\bf -k}})^T \rightarrow (\hat{a}_{2,{\bf k}},\hat{a}^{\dagger}_{2,{\bf -k}}, \hat{a}_{4,{\bf k}},\hat{a}^{\dagger}_{4,{\bf -k}})^T$ in the $k_z=\pm \pi/c$ plane. So $\omega_a=\omega_b$ and $\omega_c=\omega_d$ imply that modes $\{\hat{a}_1,\hat{a}_3\}$ decouples with $\{\hat{a}_2,\hat{a}_4\}$.  At the edge of the Brilloin zone, where operations $\Tilde{R}_{2z}$ and ${\cal T}\Tilde{M}_{2y}$ (or $\Tilde{R}_{2z}$ and ${\cal T}\Tilde{M}_{2x}$) coexist, we conclude that the four modes $(\hat{a}_{l,\bf k}, \hat{a}_{l,\bf -k})^T$ with $l=\{1,2,3,4\}$ are decoupled with eigenenergies $\{\omega_a,\omega_b,\omega_c,\omega_d\}$. Therefore, the edge of the Brillouin zone is a nodal line with fourfold degeneracy $\omega_a=\omega_b=\omega_c=\omega_d$. 

\end{appendix}

\end{document}